\documentclass[apjl]{emulateapj}

\usepackage{underscore}
\usepackage{color}
\usepackage[dvipsnames]{xcolor}

\begin{document}

\title{New observational evidence of active asteroid P/2010 A2: Slow rotation of the largest fragment}

\author{Yoonyoung Kim, Masateru Ishiguro, and Myung Gyoon Lee}
\affil{Department of Physics and Astronomy, Seoul National University,
Gwanak, Seoul 151-742, Korea\\
yoonyoung@astro.snu.ac.kr, ishiguro@astro.snu.ac.kr}

%%%%%%%%%%%%
% ABSTRACT %
%%%%%%%%%%%%

\begin{abstract}
We report new observations of the active asteroid P/2010 A2 taken when it made its closest approach to the Earth (1.06 au in 2017 January) after its first discovery in 2010. Despite a crucial role of the rotational period in clarifying its ejection mechanism, the rotational property of P/2010 A2 has not yet been studied due to the extreme faintness of this tiny object ($\sim$120~m in diameter).
Taking advantage of the best observing geometry since the discovery, we succeed in obtaining the rotational light curve of the largest fragment with Gemini/GMOS-N.
We find that (1) the largest fragment has a double-peaked period of $11.36\pm0.02$~hr spinning much slower than its critical spin period; (2) the largest fragment is a highly elongated object ($a/b\geqslant 1.94$) with an effective radius of $61.9^{+16.8}_{-9.2}$~m; (3) the size distribution of the ejecta follows a broken power law (the power indices of the cumulative size distributions of the dust and fragments are $2.5\pm0.1$ and $5.2\pm0.1$, respectively); (4) the mass ratio of the largest fragment to the total ejecta is around 0.8; and (5) the dust cloud morphology is in agreement with the anisotropic ejection model in \citet{Kim2017}. These new characteristics of the ejecta obtained in this work are favorable to the impact shattering hypothesis.

\end{abstract}
\keywords{minor planets, asteroids: individual (P/2010 A2)}

%%%%%%%%%%%%
% INTRODUCTION %
%%%%%%%%%%%%
\section{Introduction}
\label{sec:introduction}

P/2010 A2 (hereafter A2) is one of the main-belt asteroids receiving the most attention from Solar System scientists because of the mysterious dust ejection within the snow line of the Solar System \citep{Jewitt2015}. It was discovered on 2010 January 6, exhibiting a comet-like dust trail \citep{discovery}. Early analyses of its trail position angle indicated that the mass was impulsively ejected by either impact or rotational instability \citep{Jewitt2010,Snodgrass2010}.
More detailed studies through dust-modeling analysis have suggested several different mechanisms \citep[rotational breakup, impact cratering or shattering;][]{Agarwal2013,Kleyna2013,Kim2017}.
Meanwhile, the characteristics of the fragments such as the rotation, shape, and size distribution have not been studied due to the extreme faintness of this tiny object ($\sim$120~m in diameter). Specifically, the rotational period of the largest fragment (LF) is critically important in clarifying the ejection mechanism of the active asteroid \citep{Jewitt2015}.

Here, we report new observations of A2 taken when it made its closest approach to the Earth after its discovery in 2010 (i.e., the geocentric distance $\Delta$=1.06 au in 2017 January). Since small dust particles, which enclosed the fragments in the images of the early 2010 observations \citep[cf.][]{Agarwal2013}, have been swept away by solar radiation pressure over the eight years leaving behind a simple rod-shaped dust cloud, the new observation provides a more reliable data set for characterizing the fragment sizes. We utilized the golden opportunity to obtain the rotational light curve as well as the size distribution of the fragments. As a result, we have obtained the first evidence for the rotational status of the LF. In addition, we detected 10 possible sub-fragments. Based on the observational evidence, we consider the question of the cause of the mass ejection from the asteroid.

%%%%%%%%%%%%
% OBSERVATIONS %
%%%%%%%%%%%%
\section{Observations}
\label{sec:observations}
We observed A2 for two successive nights on UT 2017 January 27--28 using the 8.1~m Gemini North telescope on Mauna Kea in Hawaii, as part of the Korean priority visiting program. Images were taken with the Gemini Multi-Object Spectrograph \citep[GMOS;][]{Hook2004} with a Sloan Digital Sky Survey (SDSS) {\sl g}$'$-band filter, which is the most sensitive to the signal from A2 but less sensitive to sky background among the available filters. The image scale and the field of view are 0\rlap{.}{\arcsec}1454 pixel$^{-1}$ and $5\rlap{.}{\arcmin}5 \times 5\rlap{.}{\arcmin}5$, respectively. Through test exposures, we identified the LF by comparing two individual images but noticed that it was 2\rlap{.}{\arcmin}72 away from the position estimated by the JPL Horizons online ephemeris generator. We adjusted the position prior to the main observation. The non-sidereal rate of motion was sufficiently accurate that the telescope tracked the object adequately. The seeing was $\sim0\rlap{.}{\arcsec}8$, and the weather was photometric during the observations.
Observational data comprising a series of 300-second exposures were obtained on each night, giving 7.5~hr of total effective exposure time (Table \ref{tab:obs}).
The point-like LF and surrounding dust cloud were clearly seen even in the individual images.

%\rlap{.}{\arcsec}
%\rlap{.}{\arcmin}

\begin{deluxetable*}{cccrrccccccc}
\tablecaption{Observation summary\label{tab:obs}}
%\tablewidth{375pt}
\tabletypesize{\scriptsize}
\tablenum{1}
\tablehead{\colhead{UT Date and Time\tablenotemark{a}} & \colhead{$N$\tablenotemark{b}} & \colhead{$t$\tablenotemark{c}} & \colhead{$\nu$\tablenotemark{d}} & \colhead{$r_\mathrm{h}$\tablenotemark{e}} & \colhead{$\Delta$\tablenotemark{f}} & \colhead{$\alpha$\tablenotemark{g}} & \colhead{$\theta_{-S}$\tablenotemark{h}} & \colhead{$\theta_{-V}$\tablenotemark{i}} &  \colhead{$\delta_{\earth}$\tablenotemark{j}}}
\startdata
%2017 Jan 26 10:11-10:16 & Gemini & 2 & 600 & 28.9 & 2.033 & 1.055 & 4.34 & 122.2 & 286.6 & -1.2 \\
2017 Jan 27 10:38--13:19 & 30 & 9000 & 29.3 & 2.034 & 1.057 & 5.0 & 119.0 & 286.5 & -1.1 \\
2017 Jan 28 06:11--10:49 & 50 & 15000 & 29.6 & 2.034 & 1.059 & 5.4 & 117.0 & 286.4 & -1.0 \\
2017 Jan 28 12:18--13:09 & 10 & 3000 & 29.6 & 2.034 & 1.060 & 5.5 & 116.6 & 286.3 & -1.0
\enddata
\tablenotetext{a}{UT date and range of start times of the integrations.}
\tablenotetext{b}{Number of exposures.}
\tablenotetext{c}{Total exposure time, in seconds.}
\tablenotetext{d}{True anomaly, in degrees.}
\tablenotetext{e}{Heliocentric distance, in au.}
\tablenotetext{f}{Geocentric distance, in au.}
\tablenotetext{g}{Phase (Sun--target--observer's) angle, in degrees.}
\tablenotetext{h}{Position angle of the antisolar vector, in degrees.}
\tablenotetext{i}{Position angle of the negative heliocentric velocity vector, in degrees.}
\tablenotetext{j}{Angle of Earth above the orbital plane, in degrees.}
\end{deluxetable*}

%%%%%%%%%%%%%%%%%
%%%%%% RESULT %%%%%%
%%%%%%%%%%%%%%%%%
\section{Analysis and results}
\label{sec:result}
\subsection{Rotation, Shape, and Size of the Largest Fragment}
\label{subsec:rotation}

\begin{figure}
\epsscale{1.1}
\plotone{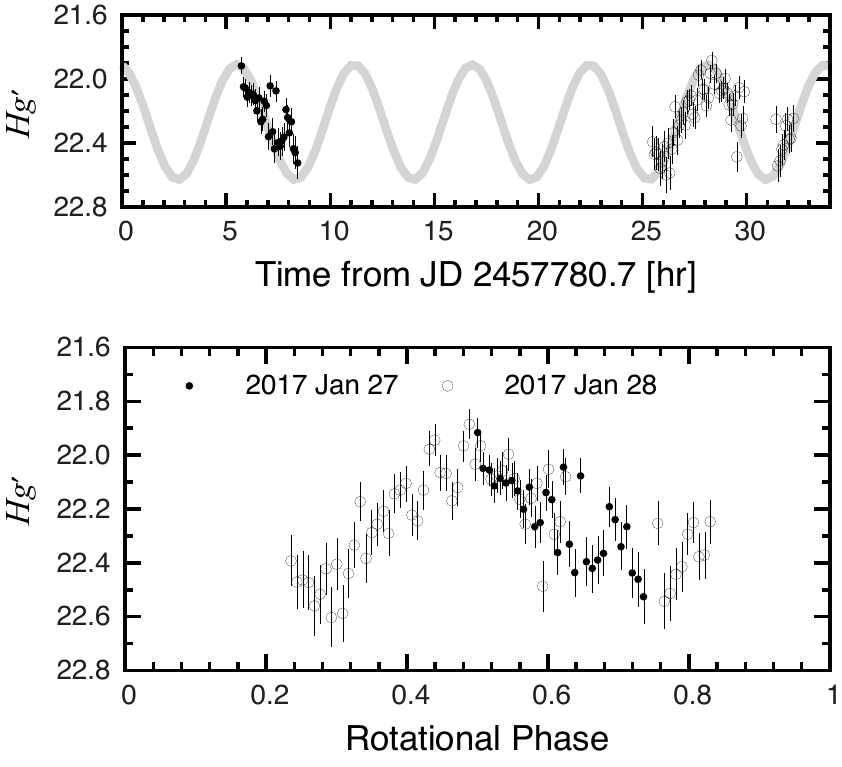}
\caption{The rotational light curve of the LF. Time-series $\textsl{g}'$-band photometry over two nights (upper panel) and phase based on the best-fit double-peaked period of 11.36 hr (lower panel). A sine curve with a period of 11.36 hr was plotted in the upper panel (gray line). Two data points were excluded whose photometry was contaminated by field stars. The bump at rotational phase 0.6--0.7 is associated with neither contaminations of background sources nor data artifacts but can be indicative of rapidly changing cross-section due to a complicated shape.
}
\label{fig:lc}
\end{figure}

We obtained photometry of the LF in each image using a circular aperture of projected radius 1\rlap{.}{\arcsec}3. The sky background was determined within a concentric annulus with projected inner and outer radii of 1\rlap{.}{\arcsec}7 and 3\rlap{.}{\arcsec}2, respectively. 
Flux calibration was performed using Landolt standard stars (PG0231+051 and PG1047+003) at similar airmass to A2, while $\sim$20 field stars in each individual image were also measured for differential photometry and used to correct for time-variable atmospheric extinction during each night.
We converted the apparent magnitudes to absolute magnitudes (i.e., the magnitude at a hypothetical point at unit heliocentric and geocentric distances and at a zero Sun--asteroid--observer's angle, the so-called phase angle) by
\begin{equation}
H_{\textsl{g}'}=m_{\textsl{g}'} - 5~\log_{10}\left(r_\mathrm{h} \Delta\right)+2.5~\log_{10}\left(\Phi\left(\alpha\right)\right)~,
\label{eq:abs}
\end{equation}

\noindent in which $r_\mathrm{h}$ is the heliocentric distance. $\Phi\left(\alpha\right)$ is the phase function at solar phase angle $\alpha$, where we used the H-G formalism with a slope parameter of $G = 0.25$ for S-complex asteroids \citep{Bowell1989}, which are dominant in the inner main-belt.
Figure \ref{fig:lc} shows the light curves of the LF measured from each image during our observations.
To determine the periodicity in the light curve, we applied the phase dispersion minimization \citep[\textit{PDM};][]{PDM} algorithm using the \textit{PDM} package in IRAF and obtained the single-peaked period of $P_0$ = 5.68$\pm$0.01 hr. Other possible periods are multiples of $P_0$, depending on the number of peaks appearing in one periodic phase.
The maximum peak-to-peak amplitude was $\Delta m = 0.72\pm0.10$, where such large modulation is expected from a double-peaked light curve caused by an elongated shape \citep{Harris2014}. Assuming that the light curve results from an rotating triaxial ellipsoidal body having the axis ratio of $a:b:c$ ($a \geqslant b \geqslant c$), we obtained a double-peaked period of $P_\mathrm{rot}=2P_0=11.36\pm0.02$~hr and a lower limit on the axis ratio of $a/b=10^{0.4\Delta m} \sim 1.94$.
Earlier determinations of the absolute magnitudes in $B$, $V$, and $R$ filters were $H_{B} = 22.77 \pm 0.02$, $H_{V} = 22.00 \pm 0.07$, and $H_{R} = 21.41 \pm 0.03$ \citep{Jewitt2010,Jewitt2013}.
We used transformation equations between the Johnson-Cousins ($UBVRI$) system and the SDSS ($\textsl{u}'\textsl{g}'\textsl{r}'\textsl{i}'\textsl{z}'$) system \citep{Smith2002}, $\textsl{g}'=V+0.54\left(B-V\right)-0.07$, and obtained the corresponding absolute magnitude in the $\textsl{g}'$ band, $H_{\textsl{g}'}=22.34\pm0.07$ for $\left(B-V\right)=0.77$, in agreement with the rotationally averaged magnitude from our photometry ($H_{\textsl{g}'} = 22.24 \pm 0.03$).
The resulting absolute magnitude is converted into the effective radius ($r_e$) by the following equation \citep{Russell1916}:

\begin{equation}
p_{\textsl{g}'} r_e^2 = 2.24\times10^{22} 10^{0.4(m_\odot-H_{\textsl{g}'})}
\label{eq:C}
\end{equation}

\noindent
where $m_\odot = -26.37$ is the apparent solar $\textsl{g}'$-band magnitude \citep{Blanton2007}.
Assuming a geometric $\textsl{g}'$-band albedo with large uncertainty, $p_{\textsl{g}'}=0.21\pm0.08$,
corresponding to the average albedo of known S-complex asteroids \citep{Usui2013,DeMeo2013},
we estimated an effective LF radius of $r_e = 61.9^{+16.8}_{-9.2}$~m.
Using the $r_e$ value, we obtained $a\sim$ 86.2~m and $b\sim$ 44.4~m.

\begin{figure*}
\epsscale{0.55}
\plotone{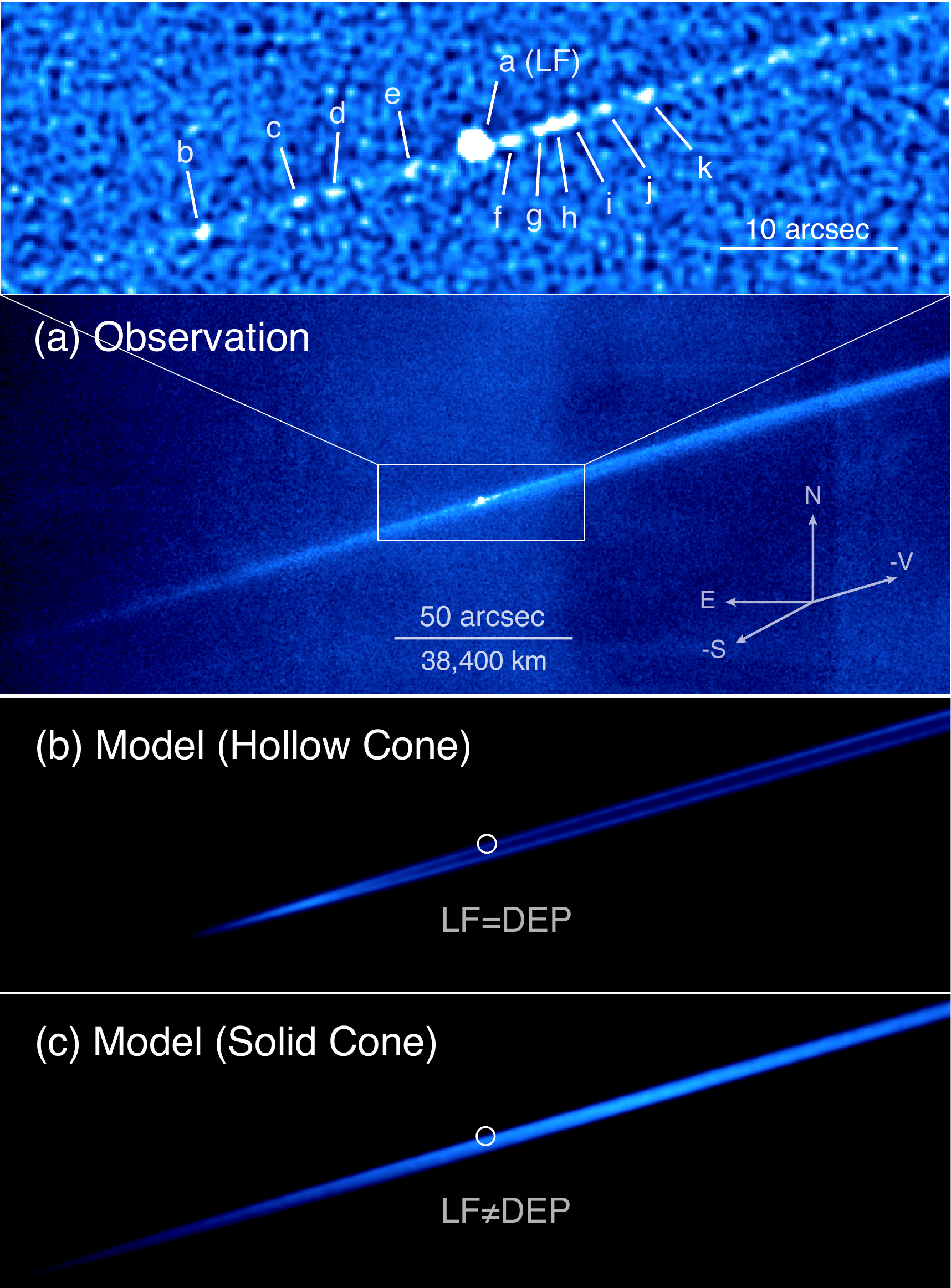}
\caption{(a) Composite image of A2 constructed from all data listed in Table \ref{tab:obs} (lower panel), where a box marks the region shown in the upper panel processed by subtracting the 11-pixel $\times$ 11-pixel ($1\rlap{.}{\arcsec}6 \times 1\rlap{.}{\arcsec}6$) median image. We applied Gaussian smoothing to enhance the visibility of the fragments. Arrows show north ($N$), east ($E$), the projected negative heliocentric velocity vector ($-V$), and the antisolar direction ($-S$).
(b) Model image of A2 assuming a hollow cone of dust as expected from an impact cratering \citep{Ishiguro2011}. The position of the LF is indicated by open circles, and the dust ejection point (DEP) is fixed to be the LF. (c) Model image of A2 assuming anisotropic ejection within a solid cone-shaped jet \citep{Kim2017}. In this model, we do not show the positions of the DEP because they exist beyond the field of view. The model images have the same image scale as the observation image, and we applied Gaussian smoothing to match the 0\rlap{.}{\arcsec}8 seeing of the data.}
\label{fig:image}
\end{figure*}

\subsection{Size Distribution of the Fragments}
\label{subsec:csd}

A composite image of A2 constructed from all data listed in Table \ref{tab:obs} is shown in Figure \ref{fig:image} (a), where background stars were removed. Several local enhancements along the trail were visible near the LF. To extract these fragments in the composite image, we first used an unsharp masking technique, subtracting the 11-pixel $\times$ 11-pixel ($1\rlap{.}{\arcsec}6 \times 1\rlap{.}{\arcsec}6$) median-filtered image. The large-scale components including the dust trail structure and sky background were subtracted by this method, leaving fine-scale structures (mostly signals from fragments). This flattened image was used for the detection of the fragments. The positions of the fragments were identified using SExtractor \citep{Bertin1996}, and each signal in the original composite image was examined with the \textit{APPHOT} package in IRAF for aperture photometry. To minimize the contaminating signals from nearby dust and fragments, we employed a small aperture of projected radius 0\rlap{.}{\arcsec}5 and obtained their magnitudes in comparison with the LF measured using the same aperture radius. Finally, we selected the point-like sources with signal-to-noise ratios greater than 3 and confirmed 10 sub-fragments as well as the LF. The fragments detected by this technique are indicated in Figure \ref{fig:image} (a).
The resulting apparent magnitudes of the sub-fragments were converted to absolute magnitudes and effective radii using Equation \ref{eq:abs} and \ref{eq:C}, respectively. For the LF, we adopted the rotationally averaged absolute magnitude and its effective radius from our light curve analysis (Section \ref{subsec:rotation}).

\begin{figure}
\epsscale{1.0}
\plotone{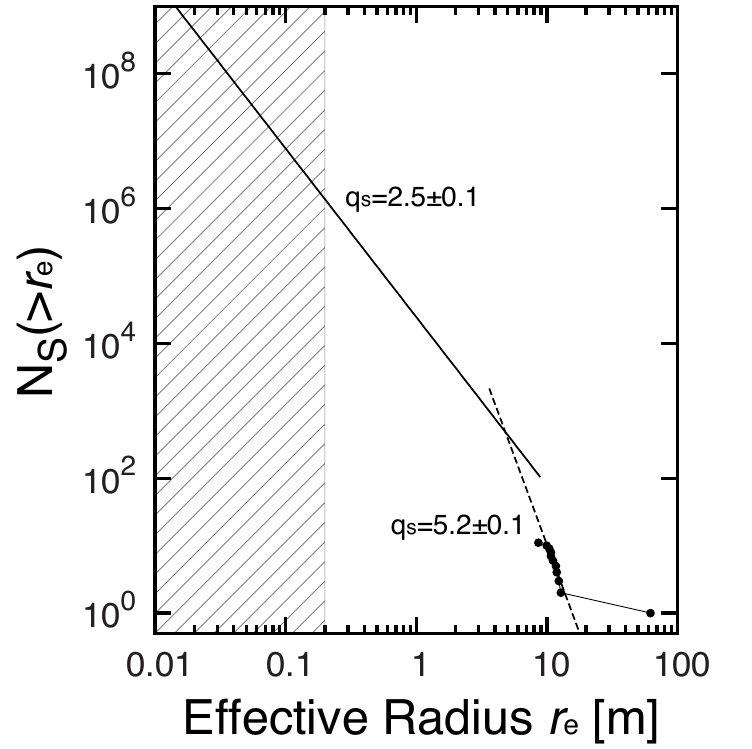}
\caption{Cumulative size distribution of the dust particles and fragments measured in this work. Filled circles denote the fragments and are fitted to the line with the power index of the cumulative size distribution, $q_S=5.2\pm0.1$ (dotted line). The shaded region denotes the dust particle regime ($\le$0.2~m), which is fitted to the line $q_S=2.5\pm0.1$ (solid line).}
\label{fig:csd}
\end{figure}

The cumulative size distribution of the fragments is shown in Figure \ref{fig:csd} (filled circles), fitted with the dashed line with the power index of the cumulative size distribution, $q_S=5.2\pm0.1$. We also show the size distribution of the dust cloud conjectured by means of a dust dynamical simulation (solid line). The power index of the dust cumulative size distribution was $q_S=2.5\pm0.1$ \citep{Jewitt2013,Kim2017}, which is significantly shallower than that of the fragments.
This discrepancy may suggest the nature of a broken power law distribution of the A2 ejecta. Similar trends are reported in the size distribution of boulders on asteroids (25143) Itokawa \citep{Michikami2008} and those of collisional asteroid families \citep{Zappala2002} but are not clear in fragmented comets like 73P and 332P \citep{Ishiguro2009,Jewitt2016}.

\bigskip
The total dust mass was determined by previous research; that is, $M_\mathrm{d}=$(5--6)$\times10^8$~kg for the sizes $<$ 0.2 m \citep{Jewitt2013}. We also obtained the consistent mass using the new observations through dust dynamical analysis (also see Section \ref{sec:discussion}).
Assuming the same mass density of the dust particles and fragments ($\rho$ = 3000 kg m$^{-3}$),
we estimated the mass of the LF and that of 10 sub-fragments as $M_\mathrm{LF}=3.0\times10^9$~kg and $M_\mathrm{sub}=1.7\times10^8$~kg, respectively. With the dust mass of $M_\mathrm{d}=6\times10^8$~kg, we obtained the mass ratio of the LF to the total ejecta (including the LF itself, 10 sub-fragments and dust), $M_\mathrm{LF}/M_\mathrm{tot}$ $\sim$ $M_\mathrm{LF}/(M_\mathrm{LF}+M_\mathrm{sub}+M_\mathrm{d})$  $\sim $0.8.

%%%%%%%%%%%%
%     DISCUSSION    %
%%%%%%%%%%%%
\section{Discussion}
\label{sec:discussion}

Here, we consider possible ejection mechanisms of A2 based on the new characteristics of the fragments obtained in this work: (i) the slow rotation of the LF ($P_\mathrm{rot}=11.36$~hr), (ii) the highly elongated shape of the LF ($a/b\geqslant 1.94$), (iii) the effective LF radius ($r_e = 61.9$~m), and (iv) the large mass ratio ($M_\mathrm{LF}/M_\mathrm{tot} = 0.8$).

We start with the possibility of rotational instability. If an elongated object is spinning beyond the critical spin period, it is likely to shed surface materials at the ends of the longest axis where the strongest centrifugal force is exerted \citep{Samarasinha2004,Hirabayashi2014}.
Applying this scenario to the case of A2, the long axis of the precursor ($a_0$) is expected to be longer than that of the resultant LF ($a_\mathrm{LF}$) after the mass ejection. Approximating the asteroid as a prolate ellipsoid whose mass is proportional to $a$ under the same mass density $(M_\mathrm{tot}/M_\mathrm{LF}=a_0/a_\mathrm{LF})$, the axis ratio of the precursor is estimated to be $\left(a/b\right)_0 = \left(M_\mathrm{tot}/M_\mathrm{LF}\right) \left(a/b\right)_\mathrm{LF} \sim 2.43$.
Such highly elongated objects ($a/b>2.4$) are extremely rare in the main asteroid belt but are present in some asteroid families \citep{Szabo2008}.

The rapid rotation of the primary body has been considered to be a key feature of rotationally disrupted asteroids. 
For a strengthless elongated body, the critical spin period for breakup (at which the gravitational acceleration equals the centripetal acceleration at the equator) is given by \citet{Jewitt2012}

\begin{equation}
P_\mathrm{crit}=\left(a/b\right)_0\left[\frac{3\pi}{G\rho}\right]^{1/2}~,
\end{equation}

\noindent where $\left(a/b\right)_0$=2.43 is the axis ratio of the precursor and $\rho$ is the mass density of the object. 
With $\rho$ = 3000 kg m$^{-3}$ for a monolithic precursor with a composition similar to S-complex asteroids, we obtained a critical spin period of $P_\mathrm{crit}=4.63$~hr. Alternatively, a slower critical spin period of $P_\mathrm{crit}=5.82$~hr can be obtained assuming a weak rubble pile precursor like (25143) Itokawa \citep[$\rho$ = 1900 kg m$^{-3}$,][]{Fujiwara2006}.

It is reasonable to think that the spin period of the LF is constant after the mass ejection occurred in 2009 because YORP spin-down \citep{Rubincam2000} or relaxation into the minimum rotational energy \citep{Burns1973,Jewitt2004} is not expected within the timescale of eight years. Although quantitative constraints on potential angular momentum loss during rotational breakup have not yet been studied (M. Hirabayashi 2017, private communication), we conjecture that it would be difficult to support the idea of the rotational breakup because the current spin period of the LF ($11.36\pm0.02$~hr) is fairly longer than the $P_\mathrm{crit}$. It is also noteworthy that other active asteroids whose ejection mechanism are presumed to be rotation-related, have revealed a rapid rotation close to their critical spin period \citep[133P, 331P, and 62412;][]{Hsieh2004,Drahus2015,Sheppard2015}.

Turning to the hypothesis of an impact cratering, we considered the potential crater size on the surface of the LF, based on the updated sizes and masses. Assuming that the LF lost its mass as the observed dust and 10 fragments through a cratering event where their mass density was the same ($\rho$ = 3000 kg m$^{-3}$), we obtained the total ejecta mass of $7.7\times10^8$~kg (cf. Section \ref{subsec:csd}), corresponding to the total ejecta volume of $2.6\times10^5$~m$^{3}$.
The ejecta volume was set to be equal to the paraboloid crater volume $V=1/3\pi R_\mathrm{c}^3$, where we assume that the transient crater depth is a third of its diameter \citep{Richardson2007}.
As a result, we obtained the crater radius of $R_\mathrm{c}=63$~m. The resulting crater radius on the surface of the LF is equivalent to  the effective LF radius ($r_e = 61.9$~m), which is unlikely \citep[cf.][]{Hainaut2012}.
In addition, we conducted a model simulation of dust particles assuming a hollow cone of dust as expected from an impact cratering. The basic algorithm for the simulation is essentially the same as that described in \citet{Ishiguro2011}, but appropriate model parameters were applied to reproduce the A2 dust cloud morphology \citep[symmetric ejection with respect to a vector normal to the asteroid surface $(\alpha_\mathrm{cone},~\delta_\mathrm{cone})=(30\arcdeg,-15\arcdeg)$ with a half-opening angle of $\theta=35\arcdeg$ was assumed; cf.][]{Kleyna2013}.
In the simulation image (Figure \ref{fig:image} (b)), the doublet structure associated with an impact cone would have been detected in the observed image. Although there are several solar system objects that have craters equivalent to the bodies' diameter \citep[e.g., Mathilde and Deimos,][]{Burchell2010}, it is less probable that the mass ejection occurred at A2 through impact cratering because of the inconsistency of the morphology.

Most recently, \citet{Kim2017} suggested that the precursor asteroid was shattered by an impact and that remnants of slow ``antipodal'' ejecta were observed as the debris cloud of A2. We performed a model simulation of the dust particles and large fragments at the epoch of 2017 Gemini observation using the same algorithm and parameters in \citet{Kim2017}. The model was found to be still suitable for our new observations, reproducing the dust cloud morphology and the trail surface brightness (Figure \ref{fig:image} (c)) as well as the positions of fragments \citep[details in][]{Kim2017}.
Additionally, we note that the slow rotation of the LF is in agreement with the impact shattering hypothesis; that is, laboratory impact experiments suggest that large or antipodal fragments tend to spin slowly \citep{Fujiwara1981,Nakamura1992}.

While it is true that the large mass ratio of the LF to the total ejecta ($M_\mathrm{LF}/M_\mathrm{tot} = 0.8$) remains inconsistent with the impact shattering hypothesis, where the smaller mass ratio ($\lesssim0.5$) is generally assumed \citep{Holsapple2002}, we strongly maintain that impact shattering is the likely mechanism of the activity of A2. Perhaps pre-existing fractures and voids in the precursor body enable the LF to be less damaged in the disruptive impact and maintain a larger mass, while a smaller $M_\mathrm{LF}/M_\mathrm{tot}$ ratio can be obtained if we assume that the mass density of the dust particles is larger than that of the LF. If we consider the mass of invisible objects in the size range of 0.2--8~m, $M_\mathrm{tot}$ would increase. Therefore, $M_\mathrm{LF}/M_\mathrm{tot}$ would be the upper limit.
Although our new observations would not constitute conclusive evidence for the activity of A2, they provide further evidence (slow rotation of the LF as well as the consistency of the anisotropic dust ejection model in \citet{Kim2017}) favorable to the hypothesis of impact shattering (i.e., catastrophic disruption).

\acknowledgments
We thank A. M. Nakamura, M. Hirabayashi, J. {\v D}urech, D. Jewitt and the anonymous referee for their valuable comments. This work was supported by the National Research Foundation of Korea (NRF) funded by the Korean government (MEST; No. 2012R1A4A1028713) and the K-GMT Science Program (PID: GN-2016B-Q-14) of the Korea Astronomy and Space Science Institute (KASI). S.-C. Yang, H.-G. Lee and priority visitors helped with the Gemini observations.

\end{document}